\documentclass[11pt]{article}
\usepackage{amsfonts}

\usepackage{amscd}
\usepackage{amsmath}
\newlength{\dinwidth}
\newlength{\dinmargin}
\def\be{\begin{equation}}
\def\ee{\end{equation}}
\def\ben{\begin{displaymath}}
\def\een{\end{displaymath}}
\def\baa{\begin{eqnarray}}
\def\eaa{\end{eqnarray}}

\def\ba{\begin{array}}
\def\ea{\end{array}}

\makeatletter
\@addtoreset{equation}{section}
\makeatother

\def\l{\lambda}

\def\CP1{{\mathbb P}^1}

\def\la{\label}

\def\f{\frac}
\def\L{{\cal L}}
\def\p{\partial}

\def\log{\ln}

\def\la{\label}

\def\f{\frac}
\def\L{{\cal L}}
\def\p{\partial}

\newtheorem{remark}{Remark}

\newtheorem{theorem}{Theorem}

\newtheorem{lemma}{Lemma}
\newtheorem{proposition}{Proposition}
\begin{document}

\title{
On $G$-function of Frobenius manifolds related to Hurwitz spaces}
\author{A. Kokotov\footnote{e-mail: alexey@mathstat.concordia.ca}
 \;  and
D. Korotkin
\footnote{e-mail: korotkin@mathstat.concordia.ca}}

\maketitle

\begin{center}
Department of Mathematics and Statistics, Concordia University\\
7141 Sherbrooke West, Montreal H4B 1R6, Quebec,  Canada
\end{center}

\vskip0.5cm

\vskip0.5cm

Running title: On $G$-function of Frobenius manifolds

\vskip0.5cm
Keywords: Frobenius manifolds, Hurwitz spaces,
isomonodromic deformations

\vskip0.5cm
AMS subject classification: 32G99
\vskip0.5cm
{\bf Abstract.}
The semisimple Frobenius manifolds related to the Hurwitz spaces $H_{g, N}(k_1, \dots, k_l)$
are considered. We show that the corresponding isomonodromic tau-function $\tau_I$
coincides with $(-1/2)$-power of the Bergmann tau-function  which was introduced in a recent work by the authors
\cite{KokKor}.
This enables us to calculate explicitly the $G$-function of Frobenius manifolds related to the Hurwitz spaces
$H_{0, N}(k_1, \dots, k_l)$ and $H_{1, N}(k_1, \dots, k_l)$. As simple consequences we get formulas
for the $G$-functions of the Frobenius manifolds ${\mathbb C}^N/\tilde{W}^k(A_{N-1})$ and
 ${\mathbb C}\times{\mathbb C}^{N-1}\times\{\Im z >0\}/J(A_{N-1})$, where $\tilde{W}^k(A_{N-1})$ is an extended affine Weyl group and
$J(A_{N-1})$ is a Jacobi group, in particular, proving the conjecture of \cite{Strachan}.
In case of Frobenius manifolds related to Hurwitz spaces $H_{g, N}(k_1, \dots, k_l)$
with $g\geq2$ we obtain formulas for $|\tau_I|^2$
which allows to compute the real part of the $G$-function.
\newpage

\section{Introduction}
In these notes we deal with the class of Frobenius manifolds related to Hurwitz spaces of moduli
of meromorphic functions on Riemann surfaces (see \cite{D}).

The key observation of the present paper is the identification of the isomonodromic tau-function
(see \cite{D}, \cite{DZ1}, \cite{DZ2}, \cite{DZ3})
of this class of Frobenius manifolds with $(-1/2)$-power of the Bergmann tau-function which was
introduced in \cite{KokKor} in rather different context.
We show that the quadratic Hamiltonian from \cite{D} coincides (up to a constant) with the value of the
Bergmann projective connection calculated in the natural local parameter at the critical point of the meromorphic
function. This simple
observation enables us to apply the results of \cite{KokKor} and explicitly calculate
the isomonodromic tau-functions of Frobenius manifolds related to the Hurwitz spaces of moduli of
meromorphic functions on surfaces of genus $0$ and $1$. This immediately leads to general formulas
for the $G$-function (see \cite{DZ2} and \cite{DZ3}) of the above Frobenius manifolds .
(We recall that
the $G$-function of a Frobenius manifold
provides a solution of the so-called Getzler equation (see \cite{Getzler}, \cite{DZ3}); for some classes of Frobenius manifolds it plays a role
of generating function of Gromov-Witten invariants of algebraic varieties (\cite{DZ2}); in the general case
it describes first-order deformations of dispersionless integrable systems.)

As a simple
consequence we prove the recent conjecture of Strachan \cite{Strachan} which claims the following formula
for the $G$-function of the Frobenius manifold
${\mathbb C}\times{\mathbb C}^{N-1}\times\{\Im z >0\}/J(A_{N-1})$:
$$G=-\log\eta(t_0)-\frac{N+1}{24}t_N.$$
Moreover, using the results of \cite{KokKor}, we get the expression for the modulus square of the
isomonodromic tau-function (and, hence, for the real part of the $G$-function) in case of Hurwitz
spaces in higher genus.

The present work was inspired by \cite{Strachan}
where an alternative approach to the calculation of the $G$-function of Frobenius manifold was developed.

\section{Preliminaries}
In this section we briefly outline some basic facts and definitions from the theory of Frobenius manifolds
(\cite{D}, \cite{DZ1}, \cite{DZ2}, \cite{DZ3}, \cite{Nat1}, \cite{Bert}, \cite{Manin}).
\subsection{Hurwitz spaces and  Frobenius manifolds}
Here we mainly follow \cite{D}, Lecture 5,  departing somewhat from Dubrovin's original notation.
Let $H_{g, N}(k_1, \dots, k_l)$ be the Hurwitz space of equivalence classes
$[p:\L\rightarrow{\mathbb P}^1]$ of $N$-fold
branched coverings
\begin{equation}\label{branch}
p:\L\rightarrow{\mathbb P}^1,
\end{equation}
where $\L$ is a compact Riemann surface of genus $g$ and the holomorphic map $p$ of degree $N$
is subject to the following conditions:
\begin{itemize}
\item it has $M$ simple ramification points $P_1, \dots, P_M\in \L$ with distinct {\it finite} images
$\l_1, \dots, \l_M\in {\mathbb C}\subset {\mathbb P}^1$,
\item the preimage $p^{-1}(\infty)$ consists of $l$ points: $p^{-1}(\infty)=\{\infty_1, \dots, \infty_l\}$,
the ramification index of the map $p$ at the point $\infty_j$ is $k_j$ ($1\leq k_j\leq N$).
\end{itemize}

(The ramification index at a point is equal to the number of sheets of the covering which are glued at
this point, a point $\infty_j$ is a ramification point if and only if $k_j>1$. A ramification point is simple
if the corresponding ramification index  equals  $2$.)

Notice that $k_1+\dots+k_l=N$ and $M=2g+l+N-2$.
(The last equality is a consequence of the Riemann-Hurwitz formula.)
Two branched coverings $p_1:\L_1\rightarrow {\mathbb P}^1$ and $p_2:\L_2\rightarrow {\mathbb P}^1$ are called
equivalent if there exists a biholomorphic map $f:\L_1\rightarrow\L_2$ such that $p_2f=p_1$.

The Hurwitz spaces $H_{g, N}(k_1, \dots, k_l)$ can be also described as
spaces of meromorphic functions of degree $N$ on Riemann surfaces
of genus $g$ with $l$ poles of orders $k_1, \dots, k_l$ and simple critical
values.

For example, the space $H_{0, N}(N)$ has an equivalent description as the space of polynomials
$\L={\mathbb P}^1\ni z\mapsto \l(z)\in{\mathbb P}^1$
\begin{equation}\label{polynom}
\l(z)=z^N+a_2z^{N-2}+a_3z^{N-3}+\dots+a_N,
\end{equation}
whereas the space $H_{0, N}(k, N-k)$ ($1\leq k\leq N-1$) can be described as the space of
''trigonometric polynomials" $\L={\mathbb P}^1\ni z\mapsto \l(z)\in{\mathbb P}^1$
\begin{equation}\label{trigpolynom}
\l(z)=z^k+b_1z^{k-1}+\dots+\frac{b_N}{z^{N-k}}; \ \ b_N\neq 0.
\end{equation}
We assume that the critical values of $\l(z)$ in (\ref{polynom}) and (\ref{trigpolynom}) are simple
(i. e. the derivative $\l'(z)$ has only simple roots and $\l(z_i)\neq\l(z_j)$ for distinct roots
$z_1, z_2$ of $\l'(z)$).

Introduce also the covering $\hat{H}_{g, N}(k_1, \dots, k_l)$ of the space $H_{g, N}(k_1, \dots, k_l)$
consisting of pairs
$$\Big<\big[p:\L\rightarrow {\mathbb P}^1\big]\in H_{g, N}(k_1, \dots, k_l)\ ,\,
\{a_\alpha, b_\alpha\}_{\alpha=1}^g\Big>,$$
where $\{a_\alpha, b_\alpha\}_{\alpha=1}^g$ is a canonical basis of cycles on the Riemann surface $\L$.

Obviously, for $g=0$ the spaces $H_{0, N}(k_1, \dots, k_l)$ and $\hat{H}_{0, N}(k_1, \dots, k_l)$ coincide.

The spaces $H_{g, N}(k_1, \dots, k_l)$ and $\hat{H}_{g, N}(k_1, \dots, k_l)$ are
connected complex manifolds of dimension $M=2g+l+N-2$, the local coordinates
on these manifold are given by the finite critical values of the map $p$ (or, equivalently, the finite branch points
of the covering (\ref{branch}))
 $\l_1, \dots, \l_M$.

In \cite{D} it was introduced the notion of  so-called ``primary" differentials on the Riemann surfaces
$\L$; each primary differential $\phi$  defines a structure of Frobenius manifold $M_\phi$
on $\hat{H}_{g, N}(k_1, \dots, k_l)$.
We will not reproduce here the complete list of primary differentials (see \cite{D}). We only notice that
in the case $g\geq1$ the normalized ($\int_{a_\alpha}\omega_\beta=
\delta_{\alpha \beta}$) holomorphic differentials $\omega_\beta$  on Riemann surfaces $\L$
are primary differentials. The (meromorphic) differentials $dz$ and $\frac{dz}{z}$ on the Riemann sphere $\L$
are primary differentials  in cases of the spaces $H_{0, N}(N)$ and $H_{0, N}(k, N-k)$ respectively.

The structure of Frobenius manifold $M_\phi$ on $\hat{H}_{g, N}(k_1, \dots, k_l)$
is defined by the multiplication law in the tangent bundle
:
$\partial_{\l_m}\circ\partial_{\l_n}=\delta_{mn}
\partial_{\l_m}$,
the unity $e=\sum_{m=1}^M\partial_{\l_m}$, the Euler field $E=\sum_{m=1}^M\l_m
\partial_{\l_m}$ and the one-form
$\Omega_{\phi^2}=\sum_{m=1}^M\Big\{{\rm Res}_{P_m}\frac{\phi^2}{d\l}\Big\}d\l_m$,
where $\l$ is the coordinate on the $\L$ lifted from the base ${\mathbb P}^1$.
The invariant metric $\eta(v, w)=\Omega_{\phi^2}(v\circ w)$
on the Frobenius manifold turns out to be flat and potential (i. e.  Egoroff-Darboux metric). In the coordinates $\l_1, \dots, \l_M$ (which are
called {\it canonical}) this metric is diagonal
\begin{equation}\label{metric}
\eta=\sum_{m=1}^M\eta_{mm}(d\l_m)^2\,;\ \ \ \eta_{mm}={\rm Res}_{P_m}\left(\frac{\phi^2}{d\l}\right)
\end{equation}
and its rotation coefficients
$\gamma_{mn}=\frac {\partial_{\l_n}\sqrt{\eta_{mm}}} {\sqrt{\eta_{nn}}}$ ($m\neq n$)
have the following properties:
First, they are independent of the choice of a primary differential $\phi$.
Second, they satisfy the equations
\begin{equation}\label{gamma1}
\partial_{\l_k}\gamma_{mn}=\gamma_{mk}\gamma_{kn}\, ,\ \ \ {\rm for\ \ distinct\ } k, n, m;
\end{equation}
\begin{equation}\label{gamma2}
e(\gamma_{mn})=\sum_{k=1}^M\partial_{\l_k}\gamma_{mn}=0,
\end{equation}
which provide the flatness of metric (\ref{metric}). Finally, the action of the Euler vector field
on $\gamma_{mn}$ has the form
\begin{equation}\label{gamma3}
E(\gamma_{mn})=\sum_{k=1}^M\l_k\partial_{\l_k}\gamma_{mn}=-\gamma_{mn}.
\end{equation}

The following three examples of Frobenius manifolds related to Hurwitz spaces
are of special interest, since they arise also in the theory of (respectively)
Coxeter, extended affine Weyl and Jacobi groups (see \cite{D}, \cite{DZ1}, \cite{Bert}).
\begin{itemize}
\item  ${\bf M_{0; N}.}$
The underlying Hurwitz space here is the space $H_{0, N}(N)$.
In this case $g=0$, $l=1$; the primary differential defining the structure
of Frobenius manifold is $dz$.
\item  ${\bf M_{0; k, N-k}.}$
The underlying Hurwitz space is   ${H}_{0, N}(k, N-k)$ ($g=0$, $l=2$,
$1\leq k\leq N-1$); the Frobenius structure is defined by the primary differential $\frac{dz}{z}$.
\item ${\bf \hat{M}_{1, N}.}$ The underlying space here is the covering $\hat{H}_{1, N}(N)$, $g=1$,
$l=1$; the primary differential on the elliptic surface $\L$ is the normalized
($\int_a\omega=1$) holomorphic differential $\omega$.
\end{itemize}

Due to \cite{DZ1}, the first $N-1$ flat coordinates of the metric $\eta$ in case of the Frobenius manifold $M_{0; k, N-k}$
of  dimension $M=N$ are given by
$$t_\mu=(-1)^{\mu+1}\frac{k}{\mu}{\rm Res}_{z=\infty}\left[\l(z)\right]^{\mu/k}d\log z, \ \ 1\leq\mu<k-1,$$
$$t_{N-\mu}=(-1)^\mu\frac{N-k}{\mu}{\rm Res}_{z=0}\left[(-1)^N\l(z)\right]^{\mu/(N-k)}d\log z, \ \ 1\leq\mu\leq N-k.$$
The last flat coordinate $t^N$ is defined by the equation
\begin{equation}\label{ploskdva}
b_N=(-1)^N\exp\left[(N-k)t_N\right].
\end{equation}

To write down the flat coordinates on the Frobenius manifold $\hat{M}_{1; N}$ (of dimension $N+1$)
set $z(P)=\int_{\infty_1}^P\omega$, where $\infty_1$ is the point on $\L$ such that
$p(\infty_1)=\infty$ and $\l(z(P))=p(P)$. Then the flat coordinates $t_0, \dots, t_{N}$ are given by
(see \cite{Bert}): $t_0=\int_b\omega=\sigma$,
where $\sigma$ is the modulus of the elliptic curve $\L$,
$t_1=\int_a\l(z(P))dz(P)$ and
\begin{equation}\label{plosktor}
t_\mu={\rm Res}_{z=0}\,z[\l(z)]^{-\frac{\mu-1}{N}}d\l(z), \ \ \mu=2, \dots, N.
\end{equation}

\subsection{Isomonodromic tau-function and $G$-function of Frobenius manifold}
Let $M_\phi$ be the Frobenius manifold with underlying Hurwitz space $\hat{H}_{g, N}(k_1, \dots, k_l)$
and the Frobenius  structure given by a primary differential $\phi$. Set $\Gamma=||\gamma_{mn}||_{m, n=1, \dots,
M}$ (the diagonal elements of the matrix $\Gamma$ are not defined),
$U={\rm diag}(\l_1, \dots, \l_M)$ and
$V=\left[\Gamma, U\right]$.
Here $\gamma_{mn}$ are the rotation coefficients of the metric (\ref{metric}), $\l_1, \dots, \l_M$ are the canonical coordinates
on $M_\phi$. The matrix $V$ is well-defined since the diagonal elements of $\Gamma$ do not enter the
commutator $[\Gamma, U]$.

The isomonodromic tau-function $\tau_I$ of the Frobenius manifold $M_\phi$ is defined by the system of (compatible)
equations
\begin{equation}\label{tauDubr}
\frac{\partial\log \tau_I}{\partial \lambda_m}=H_m;\ \ \ m=1, \dots, M,
\end{equation}
where the Hamiltonians $H_m$ are defined by
\begin{equation}\label{Hamilt}
H_m=\frac{1}{2}\sum_{n\neq m; 1\leq n\leq M}\frac{V_{nm}^2}{\l_m-\l_n}; \ \ m=1, \dots, M.
\end{equation}

Let $t_1, \dots, t_M$ be the flat coordinates on the Frobenius manifold $M_\phi$. The
Jacobian $J={\rm det}||\frac{\partial \l_m}{\partial t_n}||$
can be expressed as follows in terms of metric coefficients $\eta_{mm}$:
\begin{equation}\label{jac}
J=\left(\prod_{m=1}^M\eta_{mm}\right)^{1/2}=\left(\prod_{m=1}^M{\rm Res}_{P_m}\frac{\phi^2}{d\l}\right)^{1/2}.
\end{equation}
The $G$-function of the Frobenius manifold $M_\phi$ is defined as follows
\begin{equation}\label{G-f}
G=\log\left(\frac{\tau_I}{J^{\frac{1}{24}}}\right).
\end{equation}

\section{Isomonodromic tau-function of Frobenius manifold and Bergmann tau-function on  Hurwitz
space}
\subsection{Rotation coefficients of the flat metric $\eta$ and the Bergmann kernel}

First, we recall the definition of the Bergmann kernel.
In the case $g>0$ the Bergmann kernel on the Torelli marked Riemann surface $\L$ is defined
by $B(P, Q)=d_Pd_Q\log E(P, Q),$
where $E(P, Q)$ is the prime-form on $\L$ (see \cite{Fay}).
At the diagonal $P=Q$ the Bergmann kernel is singular:
\begin{equation}\label{funH}
B(x(P), x(Q))=\left(\frac{1}{(x(P)-x(Q))^2}+H(x(P), x(Q)\right)dx(P)\,dx(Q),
\end{equation}
where
\begin{equation}\label{Bcon}
H(x(P), x(Q))=\frac{1}{6}S_B(x(P))+o(1)
\end{equation}
as $P\to Q$. Here $x(P)$ is a local coordinate of a point $P\in\L$, $S_B$ is the Bergmann
projective connection (see, e.g., \cite{Fay},\cite{Tyurin}).

If $g=0$ and $z:\L\rightarrow{\mathbb P}^1$ is a biholomorphic map then the Bergmann
kernel is defined by
$$B(z(P), z(Q))=\frac{dz(P)dz(Q)}{(z(P)-z(Q))^2}.$$
(In particular $S_B(z)\equiv 0$ in the local parameter $z$.)

Near a simple ramification point $P_m\in \L$ of covering (\ref{branch}) we introduce the local parameter
\begin{equation}\label{loc}
x_m(P)=\left(\l(P)-\l_m\right)^{1/2},
\end{equation}
where $\l(P)=p(P)$, $\l_m=p(P_m)$; $m=1, \dots, M$.

Let $U(P_m)$ and $U(P_n)$ be small neighborhoods of ramification points $P_m$ and $P_n$. For
$(P, Q)\in U(P_m)\times U(P_n)$ we set
$$b_{mn}(P, Q)=\frac{B(x_m(P), x_n(Q))}{dx_m(P)\,dx_n(Q)}.$$
\begin{lemma}\label{L1}(cf. \cite{KokKor1})
The rotation coefficients $\gamma_{mn}$ of the metric
$\eta=\sum_{m=1}^M{\rm Res}_{P_m}\left(\frac{\phi^2}{d\l}\right)(d\l_m)^2$
are related to $b_{mn}(P, Q)$ as follows
\begin{equation}\label{gB}
\gamma_{mn}=\frac{1}{2}b_{mn}(P_m, P_n); \ \ \ m, n=1, \dots, M; \ m\neq n.
\end{equation}
\end{lemma}
{\bf Proof.} For $g\geq1$ the proof is contained in \cite{KokKor1}. In brief, it looks as follows:
Since the rotation coefficients are independent of the choice of a primary differential
$\phi$, it is sufficient to verify (\ref{gB}) only in the case $\phi=\omega_1$, where $\omega_1$ is the holomorphic differential
on $\L$ such that $\int_{a_\alpha}\omega_1=\delta_{1\alpha}$. For such a primary differential we have
$$\eta_{mm}={\rm Res}_{P_m}\frac{\omega_1^2}{d\l}=\frac{1}{2}
\left[\frac{\omega_1(x_m(P))}{dx_m(P)}\Big|_{P=P_m}\right]^2.$$
Now (\ref{gB}) follows from the definition of rotation coefficients and the Rauch formula:
\begin{equation}\label{Rauch}
\frac{\partial}{\partial \l_n}\left[\frac{\omega_1(x_m(P))}{dx_m(P)}\Big|_{P=P_m}\right]=
\frac{1}{2}b_{mn}(P_m, P_n)\left[\frac{\omega_1(x_n(P))}{dx_n(P)}\Big|_{P=P_n}\right].
\end{equation}
Consider the case $g=0$. Let $z:\L\rightarrow {\mathbb P}^1$ be a biholomorphic map such that $z(\infty_1)=\infty$.
Then $\phi=dz$ is a primary differential in the sense of Dubrovin. For this primary differential
\begin{equation}\label{rod0}
\eta_{mm}={\rm Res}_{x_m=0}\frac{[z'(x_m)dx_m]^2}{2x_mdx_m}=\frac{1}{2}\Big\{z'(x_m)\Big|_{x_m=0}\Big\}^2.
\end{equation}
Let us prove an analog of Rauch's variational formula for the meromorphic differential $dz$.
Setting $\alpha_m=z'(x_m)\Big|_{x_m=0}$, we get
\begin{equation}\label{as1}
\frac{\partial}{\partial \l_n}\{dz\}=\frac{\partial}{\partial \l_n}\left[\big(\alpha_m+O(\sqrt{\l-\l_m})\big)
\frac{d\l}{2\sqrt{\l-\l_m}}\right]=\left(\frac{\delta_{mn}\alpha_m}{2x_m^2}+O(1)\right)dx_m
\end{equation}
as $x_m\to 0$. Thus, the meromorphic differential $\frac{\partial}{\partial \l_n}dz$ has the only pole at
$P_n$ and, therefore,
\begin{equation}\label{diff}
\frac{\partial}{\partial \l_n}\{dz(P)\}=\frac{1}{2}\left[\frac{B(P, x_n)z'(x_n)}{dx_n}\Big|_{x_n=0}\right].
\end{equation}
On the other hand as $P\to P_m$ for $m\neq n$, we have
$$\frac{\partial}{\partial \l_n}\{dz\}=\frac{\partial}{\partial \l_n}(\alpha_m+O(x_m))dx_m=
\left(\frac{\partial \alpha_m}{\partial \l_n}+O(x_m)\right)dx_m.$$
Thus, due to (\ref{diff}), we get the following analog of the Rauch formula (\ref{Rauch}):
\begin{equation}\label{berg0}
\frac{\partial \alpha_m}{\partial \l_n}=\frac{\frac{\partial}{\partial \l_n}dz(x_m)}{dx_m}\Big|_{x_m=0}=
\frac{1}{2}b_{mn}(P_m, P_n)\alpha_n.
\end{equation}

Now (\ref{gB}) follows from (\ref{rod0}), (\ref{berg0}) and the definition of rotation coefficients.
$\square$
\begin{remark}\label{rem1}
{\rm Lemma \ref{L1} clarifies properties (\ref{gamma1}) -- (\ref{gamma3}) of the rotation coefficients.
Namely,  property (\ref{gamma1}) is nothing but the Rauch variational formula for the Bergmann kernel,
equations (\ref{gamma2}) and (\ref{gamma3}) follow from the invariance of the Bergmann kernel under the
translations $\l\mapsto \l+\epsilon$ and (respectively) dilatations $\l\mapsto (1+\delta)\l$ of every sheet of the covering
(\ref{branch}).
}
\end{remark}
\subsection{The Bergmann tau-function}
Introduce the quantities
$${\cal B}_{m}=-\frac{1}{12}S_B(x_m)\Big|_{x_m=0}; \ \ m=1, \dots, M,$$
where $S_B$ is the Bergmann projective connection from (\ref{Bcon}), $x_m$ is, as usually, the
local parameter (\ref{loc}) near the ramification point $P_m$.
In \cite{KokKor} it was introduced the so-called Bergmann tau-function $\tau_B$ on the Hurwitz space
$\hat{H}_{g, N}$ which is defined
by the system of equations:
\begin{equation}\label{tauB}
\frac{\partial \log \tau_B}{\partial \l_m}={\cal B}_m; \hskip0.7cm  m=1, \dots, M.
\end{equation}

The local solvability of system (\ref{tauB}) can be obtained, in particular, from the symmetry
of the Bergmann kernel and the first statement  of the following lemma.
\begin{lemma}\label{L2} The quantities ${\cal B}_m$ satisfy  the following equations:
\begin{equation}\label{b1}
\partial_{\l_n}{\cal B}_m=-\frac{1}{4}b_{mn}^2(P_m, P_n), \ \ \ m\neq m,
\end{equation}
\begin{equation}\label{b2}
e({\cal B}_m)=\sum_{n=1}^M\partial_{\l_n}{\cal B}_m=0,
\end{equation}
\begin{equation}\label{b3}
E({\cal B}_m)=\sum_{n=1}^M\l_n\partial_{\l_n}{\cal B}_m=-{\cal B}_m.
\end{equation}
\end{lemma}
{\bf Proof.}
Since  the singular part of the  Bergmann kernel in a neighborhood of
the ramification point
 $P_m$ is independent of $\{\l_n\}$, we have
\begin{equation}
\p_{\l_n} {\cal B}_m=-\f{1}{2}\left\{\p_{\l_n}
b_{mm}(P,Q)\right\}\Big|_{P=Q=P_m}\;.
\la{Bb}
\end{equation}
Computing the r.h.s. of (\ref{Bb}) via the   Rauch formula for the
Bergmann kernel:
\begin{equation}
\p_{\l_m} b_{nk}(P,Q)=\f{1}{2} b_{nm}(P,P_m) b_{mk} (P_m,Q)\;,
\la{RauchBerg}
\end{equation}
we get (\ref{b1}).

Under the translation
$\l\mapsto\l+\epsilon$
and the dilatation
$\l\mapsto(1+\delta)\l$
of each sheet of covering (\ref{branch}) (the both transformations generate  conformal isomorphisms of $\L$)
the Bergmann kernel remains invariant:
$$B^{\epsilon}(P^\epsilon, Q^\epsilon)=B^{\delta}(P^{\delta}, Q^{\delta})=B(P, Q).$$
We have the following transformation rules for the local parameter $x_m$ and the critical values $\l_m$:
$$x_m^{\epsilon}(P^{\epsilon})=x_m(P),\ \ x_m^{\delta}(P^{\delta})=(1+\delta)^{1/2}x_m(P);\ \
\l_m^{\epsilon}=\l_m+{\epsilon},\ \ \l_m^{\delta}=(1+\delta)\l_m.$$
Therefore, the function $H$ from (\ref{funH}) transforms as follows:
\begin{equation}\label{tran}
H^{\epsilon}(x_m^{\epsilon}(P^{\epsilon}), x_m^{\epsilon}(Q^{\epsilon}))=H(x_m(P), x_m(Q))
\end{equation}
and
\begin{equation}\label{dil}
H^{\delta}(x_m^{\delta}(P^{\delta}), x_m^{\delta}(Q^{\delta}))
=\frac{1}{1+{\delta}}H(x_m(P), x_m(Q)).
\end{equation}
Differentiating equations (\ref{tran}) and (\ref{dil}) with respect to $\epsilon$ and $\delta$
respectively, we get
\begin{equation}\label{de}
\frac{dH^{\epsilon}}{d{\epsilon}}=\sum_n\frac{\partial H^{\epsilon}}{\partial \l_n^{\epsilon}}=0
\end{equation}
and
\begin{equation}\label{dt}
\frac{dH^{\delta}}{d\delta}=\sum_n\l_n\frac{\partial H^{\delta}}{\partial \l_n^{\delta}}
=-\frac{1}{(1+{\delta})^2}H.
\end{equation}
Setting in (\ref{de}) and (\ref{dt}) $\epsilon=0$ and $\delta=0$ and, then, $P=Q$, we get
(\ref{b2}) and (\ref{b3}). $\square$
\subsection{Relation between $\tau_B$ and $\tau_I$}
The following simple observation provides a basis of this work.
\begin{proposition}\label{rel}
The Bergmann tau-function $\tau_B$ from \cite{KokKor} and the isomonodromic tau-function
$\tau_I$ are related as follows
\begin{equation}\label{ttrel}
\tau_I=(\tau_B)^{-1/2}.
\end{equation}
\end{proposition}
{\bf Proof.} Let $H_m$ be the quadratic Hamiltonians from (\ref{Hamilt}). Due to lemmas \ref{L1} and \ref{L2},
we have
$$H_m=\frac{1}{2}\sum_{n\neq m}\frac{V_{mn}^2}{\l_m-\l_n}=\frac{1}{2}\sum_{n\neq m}\gamma_{mn}^2(\l_m-\l_n)=$$
$$=\frac{1}{8}\sum_{n\neq m}b_{mn}^2(P_m, P_n)(\l_m-\l_n)=-\frac{1}{2}\sum_{n\neq m}(\l_m-\l_n)\partial_{\l_n}
{\cal B}_m=-\frac{1}{2}\left(\l_m\sum_{n\neq m}\partial_{\l_n}{\cal B}_m-\sum_{n\neq m}\l_n\partial_{\l_n}
{\cal B}_m\right)=$$ $$=\frac{1}{2}\sum_{n=1}^M\l_n\partial_{\l_n}{\cal B}_m=-\frac{1}{2}{\cal B}_m,$$
which proves (\ref{ttrel}). $\square$
\subsection{The Bergmann tau-function for coverings with arbitrary branching over the point at infinity}
In \cite{KokKor} the Bergmann tau-function $\tau_B$ was explicitly calculated in cases of Hurwitz spaces
$H_{0, N}(1, \dots, 1)$ and $\hat{H}_{1, N}(1, \dots, 1)$. In higher genera
(i. e. for the spaces $\hat{H}_{g, N}(1, \dots, 1)$ with $g\geq 2$) in \cite{KokKor} there were found
expressions for the modulus square  $|\tau_B|^2$.
(It should be noted that in \cite{KokKor} the general situation of Hurwitz spaces of coverings with
higher multiplicities of the finite branch points was investigated. This general
case corresponds to {\it nonsemisimple}
Frobenius manifolds which are not considered here.)

A slight modification of the proofs from \cite{KokKor} leads to the explicit formulas for the Bergmann
tau-function for the Hurwitz spaces $H_{0, N}(k_1, \dots, k_l)$
and $\hat{H}_{1, N}(k_1, \dots, k_l)$
of coverings with the branching
of type $(k_1, \dots, k_l)$ over the point at infinity. (Coverings from $H_{g, N}(1, \dots, 1)$
considered in \cite{KokKor} have no branching over the point at infinity.)

First, consider the case $g=0$. Let $[p:\L\rightarrow{\mathbb P}^1]\in H_{0, N}(k_1, \dots, k_l)$. Let also
$z:\L\rightarrow {\mathbb P}^1$ be a biholomorphic map such that $z(\infty_1)=\infty$ and
\begin{equation}\label{alph}
z(P)=[\l(P))]^{1/k_1}+O(1),
\end{equation}
as $P\to\infty_1$, where $\l(P)=p(P)$.

Introduce the local parameter $\zeta_s$ near the point $\infty_s$ with $s\geq2$:
$$\zeta_s(P)=\l^{-1/k_s}(P).$$
The map $z$ near the point $\infty_s$ ($s\geq 2$) is a holomorphic function of $\zeta_s$. Near
the simple ramification point $P_m$ the map $z$ is a holomorphic function of the local parameter
$x_m$ from (\ref{loc}).
The next statement is a modification of Theorem 6 from \cite{KokKor}. Its proof is essentially the same.
\begin{proposition}\label{pro1}
The Bergmann tau-function on the Hurwitz space $H_{0, N}(k_1, \dots,
k_l)$ is given by the following expression
\begin{equation}\label{otv1}
\tau_B=\left\{\frac{\prod_{s=2}^l\left(\frac{dz}{d\zeta_s}\Big|_{\zeta_s=0}\right)^{k_s+1}
}
{\prod_{m=1}^M\frac{dz}{dx_m}\Big|_{x_m=0}}\right\}^{\frac{1}{12}}.
\end{equation}
\end{proposition}

Let now $g=1$ and $[p:\L\rightarrow{\mathbb P}^1]\in \hat{H}_{1, N}(k_1, \dots, k_l)$,
where $\L$ is an elliptic Riemann surface.
Let $\omega$ be a holomorphic (not necessarily normalized) differential on $\L$.
Introduce the notation
$$h_s=\frac{\omega(\zeta_s(P))}{d\zeta_s(P)}\Big|_{P=\infty_s}; \ \ s=1, \dots, l$$
and
$$f_m=\frac{\omega(x_m(P))}{dx_m(P)}\Big|_{P=P_m}; \ \ m=1, \dots, M.$$
Let $\sigma$ be the modulus of the  elliptic
surface $\L$. Define the Dedekind eta-function by
$$\eta(\sigma)=\left\{\frac{d}{dz}\Theta\left[^{1/2}_{1/2}\right](z, \sigma)\Big|_{z=0}\right\}^{1/3}.$$
The next statement is a modification of Theorem 5 from \cite{KokKor}.
\begin{proposition}\label{pro2}
The Bergmann tau-function on the Hurwitz space $\hat{H}_{1, N}(k_1, \dots,
k_l)$ is given by
\begin{equation}\label{otv2}
\tau_B=\eta^2\left\{\frac{\prod_{s=1}^{l}h_s^{k_s+1}}
{\prod_{m=1}^Mf_m}\right\}^{\frac{1}{12}}.
\end{equation}
\end{proposition}
Due to Riemann-Hurwitz formula the r. h. s. of (\ref{otv2}) is independent of normalization
of the holomorphic differential $\omega$.

\begin{remark}
{\rm The way to obtain (\ref{otv1}) and (\ref{otv2}) in \cite{KokKor} was somehow indirect.
Namely, these formulas were deduced from the study of
the appropriately regularized Dirichlet integral ${\mathbb S}=
\frac{1}{2\pi}\int_{\L}|\phi_\l|^2$, where $e^\phi |d\l|^2$ is
the flat metric on $\L$ obtained by projecting down the
standard metric $|dz|^2$ on the  universal covering
$\tilde\L$. The derivatives of ${\mathbb S}$ with respect to the
branch points can be expressed through the values of the
Schwarzian connection at the branch points;
 this reveals a close link of ${\mathbb S}$ with the modulus of
the Bergmann tau-function. On the other hand, the integral ${\mathbb S}$ admits
an explicit calculation via the asymptotics of the flat metric
near the branch points and the infinities of the sheets of the
covering. Moreover, it admits a ``holomorphic factorization''
i.e. it can be explicitly represented as the modulus square of some
holomorphic function, which allows one to compute the Bergmann tau-function
itself.

To prove relations (\ref{otv1}) and (\ref{otv2}) directly (i. e.  without any use of
Dirichlet integrals) remains an open problem.
}
\end{remark}

\section{$G$-function of Frobenius manifolds related to Hurwitz spaces in genera $0$ and $1$}
\subsection{The general formulas for the $G$-function}
The following two theorems are immediate consequences of propositions \ref{pro1}, \ref{pro2} and
\ref{rel}.
\begin{theorem}\label{main0}
The $G$-function of the Frobenius manifold with underlying Hurwitz space $H_{0, N}(k_1, \dots, k_l)$
and the Frobenius structure given by a primary differential $\phi$ can be expressed as follows:
\begin{equation}\label{g0}
G=\frac{1}{24}\log\left\{
\frac{\prod_{m=1}^M\frac{dz}{dx_m}\Big|_{x_m=0}}
{\prod_{s=2}^l\left(\frac{dz}{d\zeta_s}\Big|_{\zeta_s=0}\right)^{k_s+1}
\left(\prod_{m=1}^M{\rm Res}_{P_m}\frac{\phi^2}{d\l}\right)^{\frac{1}{2}}}
\right\}\, .
\end{equation}
\end{theorem}

\begin{theorem}\label{main1}
The $G$-function of the Frobenius manifold with underlying Hurwitz space $\hat{H}_{1, N}(k_1, \dots, k_l)$
and the Frobenius structure given by a primary differential $\phi$ can be expressed as follows:
\begin{equation}\label{g1}
G=\frac{1}{24}\log\left\{
\frac{\prod_{m=1}^Mf_m}
{\prod_{s=1}^lh_s^{k_s+1}\left(\prod_{m=1}^M{\rm Res}_{P_m}\frac{\phi^2}{d\l}\right)^{\frac{1}{2}}}
\right\}-\log\eta(\sigma)\, .
\end{equation}
\end{theorem}
\subsection{Examples}
\subsubsection{${\bf M_{0; N}}$}
The Frobenius manifold $M_{0; N}$ is isomorphic to the orbit space ${\mathbb C}^{N-1}/A_{N-1}$
of the Coxeter group $A_{N-1}$ (see \cite{D}).

In this case $l=1$ and the first factor at the denominator of (\ref{g0}) is absent. As a map $z$
we can take one given by (\ref{polynom}), so
$$\left(\prod_{m=1}^M{\rm Res}_{P_m}\frac{\phi^2}{d\l}\right)^{\frac{1}{48}}=
\left(\prod_{m=1}^M{\rm Res}_{x_m=0}\frac{[z'(x_m)]^2dx_m}{2x_m}\right)^{\frac{1}{48}}=
{\rm const}
\left(\prod_{m=1}^M\frac{dz}{dx_m}\Big|_{x_m=0}
\right)^{\frac{1}{24}}$$
and, therefore, $G={\rm const}$.
\subsubsection{${\bf M_{0; k, N-k}}$}
According to \cite{DZ1}, the Frobenius manifold $M_{0; k, N-k}$ is isomorphic to the orbit space
of the extended affine Weyl group $\tilde{W}^k(A_{N-1})$.

In this case $l=2$, $z$ is given by (\ref{trigpolynom}).
Using the equality $\zeta_2=\l^{-\frac{1}{N-k}}$, we get
$$\frac{dz}{d\zeta_2}=\frac{1}{\l'_z}\frac{d\l}{d\zeta_2}=
\frac{1}{b_N+O(z)}\left(\frac{z}{\zeta_2}\right)^{N-k+1}=\frac{1}{b_N+O(z)}
[b_N+O(z)]^{\frac{N-k+1}{N-k}}$$
as $z\to 0$ and
$$\frac{dz}{d\zeta_2}\Big|_{\zeta_2=0}=b_N^{\frac{1}{N-k}};\ \ \ \
\left(\frac{dz}{d\zeta_2}\Big|_{\zeta_2=0}\right)^{\frac{k_2+1}{24}}=
b_N^{\frac{N-k+1}{24(N-k)}}.$$
Since $\phi=\frac{dz}{z}$,
we have
$$\left(\prod_{m=1}^N{\rm Res}_{P_m}\frac{\phi^2}{d\l}\right)^{\frac{1}{48}}=
\left(\prod_{m=1}^N{\rm Res}_{x_m=0}\frac{[z'(x_m)]^2dx_m}{2x_m[z(x_m)]^2}\right)^{\frac{1}{48}}=
{\rm const}
\frac
{
\left(\prod_{m=1}^M
\frac{dz}{dx_m}
\Big|_{x_m=0}
\right)^
{\frac{1}{24}}
}
{\left(\prod_{m=1}^M\gamma_m\right)^{\frac{1}{24}} },$$
where $\gamma_m=z(P_m)$ are the critical points of the map $\l(z)$.

On the other hand, $M=2g+l+N-2=N$ and
$$\l'(z)=kz^{k-1}+\dots+\frac{(k-N)b_N}{z^{N-k+1}}=\frac{k\prod_{n=1}^{N}(z-\gamma_n)}{z^{N-k+1}}.$$
Therefore,
$$k\prod_{m=1}^{M}\gamma_m=\pm(N-k)b_N$$
and (up to a constant independent of $\{\l_k\}$)
$$G=-\frac{1}{24}\frac{\log b_N}{N-k}=-\frac{1}{24}t_N,$$
in agreement with the main result of \cite{Strachan}.
\subsubsection{${\bf \hat{M}_{1, N}}$}
The Frobenius manifold $\hat{M}_{1, N}$ is isomorphic to the orbit space
${\mathbb C}\times{\mathbb C}^{N-1}\times\{\Im z >0\}/J(A_{N-1})$ of the Jacobi group
$J(A_{N-1})$ (see \cite{D} and \cite{Bert}).

In this case $l=1$, $M=N+1$. Following \cite{Bert}, we start
 the enumeration of the flat coordinates  from $0$.
We have
$$\left(\prod_{m=1}^M{\rm Res}_{P_m}\frac{\phi^2}{d\l}\right)^{\frac{1}{48}}=
{\rm const}\left(\prod_{m=1}^M\frac{\omega(x_m(P))}{dx_m(P)}\Big|_{P=P_m}\right)^{\frac{1}{24}}
=\left(\prod_{m=1}^Mf_m\right)^{\frac{1}{24}}.$$
On the other hand, since $\zeta_1=\l^{-\frac{1}{N}}$,
$$t_N={\rm Res}_{z=0}\big(z[\l(z)]^{-\frac{N-1}{N}}d\l(z)\big)={\rm Res}_{\zeta_1=0}\left(z(\zeta_1)\frac{d\zeta_1}
{\zeta_1^2}\right)=z'(\zeta_1)\big|_{\zeta_1=0}=h_1$$
and
$$G=-\log\eta(t_0)-\frac{N+1}{24}t_N,$$
which proves the conjecture from \cite{Strachan}.
\section{Some remarks on higher genus case}
Here we give a formula for the modulus square of the
tau-function of Frobenius manifolds related to the Hurwitz spaces
$H_{g, N}(k_1, \dots, k_l)$ with $g\geq2$. From this formula one can derive an expression for the real
part of the corresponding $G$-function.
For simplicity, we consider only the case $k_1=\dots=k_l=1$,
the results in the general case differ insignificantly.

If the covering $\L$ has genus $g>1$ then it
is biholomorphically equivalent to the quotient space ${\mathbb H}/\Gamma$, where
${\mathbb H}=\{z\in {\mathbb C}\,:\, \Im z>0\}$; $\Gamma$ is a strictly hyperbolic  Fuchsian group.
Denote by $\pi_F:{\mathbb H}\rightarrow \L$ the natural projection.
Let
$x$ be a local parameter on $\L$. Introduce the standard  metric  of the constant
curvature $-1$ on $\L$:
\begin{equation}\label{locmetr}
e^{\chi(x, \bar x)}|dx|^2=\frac{|dz|^2}{|\Im z|^2}\;,
\end{equation}
where $z\in {\mathbb H}$, $\pi_F(z)=P$, $x=x(P)$.

Denote by
 $\zeta=1/\lambda$  the local coordinate in a neighborhood of the infinity of any sheet
of covering $\L$.
Introduce functions $\chi^{ext}(\l, \bar\l)$, $\chi^{int}(x_m, \bar x_m)$, $m=1, \dots, M$ and
$\chi_k^{\infty}(\zeta, \bar\zeta)$, $k=1, \dots, N$ by specifying  $x=\l$, $x=x_m$
and $x=\zeta$ (in a neighborhood of the point at infinity of $k$th sheet) in (\ref{locmetr}) respectively.

Consider the following domain on $k$th sheet of
$\L$: $\L_\rho^k=\{\l\in \L^{k}\ : \ \forall m\ \ \ |\l-\l_m|>\rho\  \& \ |\l|<1/\rho\}$,
where $\l_m$ are all the branch points which belong to the $k$-th sheet $\L^{k}$ of the covering $\L$.
(The sheet $\L^{k}$ can be considered as a copy of the Riemann sphere ${\mathbb P}^1$ with
appropriate cuts between the branch points; the domain
$\L_\rho^k$ is obtained from $\L^{k}$ by deleting small discs
around branch points belonging to this sheet, and the disc around infinity.)

The function $\chi_k^{ext}: \L^{k}\rightarrow {\mathbb R}$
is smooth in the  domain $\L_\rho^k$ for any sufficiently small $\rho>0$.
This function has finite limits at the cuts (except the endpoints which are the branch points);
at the branch points and at the infinity there are the following asymptotics
\begin{equation}\label{asy1}
|\partial_\l\chi_k^{ext}(\l, \bar\l)|^2=\frac{1}{4}|\l-\l_m|^{-2}+O(|\l-\l_m|^{-3/2})
\end{equation}
as $\l\to\l_m$ and
\begin{equation}\label{asy2}
|\partial_\l\chi_k^{ext}(\l, \bar\l)|^2=4|\l|^{-2}+O(|\l|^{-3})\;
\end{equation}
as $\l\to\infty$.
We define
the ``truncated" integral $T_\rho$ by
\begin{equation}\label{qu-qu}
T_\rho=\sum_{k=1}^N\int_{\L_\rho^k}|\partial_\l\chi_k^{ext}|^2\,|d\l\wedge d\bar\l|/2\;.
\end{equation}
Then there exists the finite limit
\begin{equation}\label{reg2}
{\rm reg}\int_\L(|\chi_\l|^2+e^\chi)\,|d\l\wedge d\bar\l|/2=\lim_{\rho\to 0}
\left(T_\rho+\sum_{k=1}^N\int_{\L^{k}}e^{\chi_k^{ext}}\,|d\l\wedge d\bar\l|/2
+(8N+4M)\pi\log \rho\right).
\end{equation}
Define the function ${\mathbb S}_F$ by
\begin{equation}\label{Main2}
{\mathbb S}_F(\l_1, \dots, \l_M)=
\frac{1}{24\pi}\left\{{\rm reg}\int_\L(|\chi_\l|^2+e^\chi)\,|d\l\wedge d\bar\l|/2\right\}+
\frac{1}{3}\sum_{n=1}^M\chi^{int}(x_n)\Big|_{x_n=0}-\frac{1}{3}\sum_{k=1}^N\chi_k^\infty(\zeta)\Big|_{\zeta=0}\; ;
\end{equation}
and introduce the  determinant of Laplacian operator (in the Poincar\'e metric)
${\rm det}\Delta=\exp\{-\zeta'(0)\}$,
where $\zeta(s)$ is the zeta-function of the Laplacian on the Riemann surface $\L$.

Let ${\mathbb B}$ be the matrix of $b$-periods of the Riemann surface $\L$.

The following theorem is a consequence of Theorem 9 from \cite{KokKor} and Lemma \ref{rel}.

\begin{theorem}\label{main2}Let $g\geq2$.
 The modulus square of the isomonodromic tau-function on $\hat{H}_{g, N}(1, \dots, 1)$
has the following representation
\begin{equation}\label{otvet1}
|\tau_I|^2=e^{{\mathbb S}_F}\frac{({\rm det}\,\Im\,{\mathbb B})^{1/2}}{({\rm det}\Delta)^{1/2}}\ .
\end{equation}
\end{theorem}
\begin{remark}
{\rm At the moment we don't know the explicit holomorphic factorization (similar to that in genera $0$
and $1$) of the r. h. s. of (\ref{otvet1}). Finding of such a factorization seems to be of great interest.
}
\end{remark}

Let $M_\phi$ be  the Frobenius manifold with underlying Hurwitz space $\hat{H}_{g, N}(1, \dots, 1)$
and the Frobenius structure given by a primary differential $\phi$.
From Theorem \ref{main2} it follows that
 the real part of the $G$-function of $M_\phi$ is given by
\begin{equation}\label{gfin}
{\rm Re}\, G=\frac{1}{2}{\mathbb S}_F+\frac{1}{4}\log
\frac{({\rm det}\,\Im\,{\mathbb B})}{({\rm det}\Delta)}-\frac{1}{48}\log\Big|\prod_{m=1}^M{\rm Res}_{P_m}
\frac{\phi^2}{d\l}\Big|\,.
\end{equation}

{\bf Acknowledgements.} We would like to thank Marco Bertola for discussions.

This work was partially supported by the grant of
Fonds pour la Formation de Chercheurs et l'Aide a la Recherche de
Quebec, the grant of Natural Sciences and Engineering Research Council
of Canada and Faculty Research Development Program of Concordia University.


\begin{thebibliography}{99}
\bibitem{Bert}Bertola M., Frobenius manifold structure on orbit space of Jacobi groups; Parts I and II,
Diff. Geom. Appl., 13 (2000), 19-41 and 13 (2000), 213-23

\bibitem{D} Dubrovin B., Geometry of 2D topological field theories, Lect. Notes Math., vol. 1620,
Springer-Verlag, Berlin, 1996, 120--348

\bibitem{DZ1} Dubrovin B., Zhang Y., Extended affine Weyl groups and Frobenius manifolds,
Comp. Math., 111 (1998) 167-219

\bibitem{DZ2}Dubrovin B., Zhang Y., Bihamiltonian hierarchies in 2D topological field theory at
one-loop approximation, Commun. Math. Phys., 198 (1998), 311-361

\bibitem{DZ3}Dubrovin B., Zhang Y., Frobenius manifolds and Virasoro constraints, Selecta Math. (N. S.)
5 (1999), 423--466

\bibitem{Fay}Fay, John D., Theta-functions on Riemann surfaces, Lect.Notes in
Math.,  352, Springer (1973)
\bibitem{Getzler} Getzler E., The jet-space of a Frobenius manifold and higher-genus Gromov-Witten invariants,
arXiv: math.AG/0211338

\bibitem{KokKor}
Kokotov A., Korotkin D., ``Tau-functions on Hurwitz spaces'', to be published
in ``Mathematical Physics, Analysis and Geometry''; math-ph 0202034
\bibitem{KokKor1} Kokotov A., Korotkin D., Some integrable systems on Hurwitz spaces, math-ph/0112051

\bibitem{Manin} Manin Yu. I., Frobenius manifolds, quantum cohomology, and moduli spaces,
AMS, 1999

\bibitem{Nat1}Natanzon S., Hurwitz spaces, Topics on Riemann surfaces and Fuchsian groups,
Madrid, 1998, London Math. Soc. Lecture Note Ser., 287, 165-177

\bibitem{Rauch50}
Rauch, H.E. Weierstrass points, branch points, and moduli of
Riemann surfaces, Comm. Pure Appl. Math. {\bf 12} 543-560  (1959)

\bibitem{Strachan} Strachan I. A. B., Symmetries and solutions of Getzler's equation
for Coxeter and extended affine Weyl Frobenius manifold, math-ph/0205012

\bibitem{Tyurin}Tyurin, A.N., Periods of quadratic differentials
(Russian), Uspekhi Mat. Nauk {\bf 33} , no. 6(204), 149-195 (1978)

\end{thebibliography}
\end{document}